\documentstyle[prb,epsfig,preprint,aps]{revtex}
\begin{document}
\draft
\title{Evidence for strong electron--phonon coupling and polarons in the 
optical response of La$_{2-x}$Sr$_x$CuO$_4$}
\author{Oleg V Dolgov}
\address{P. N. Lebedev Physical Institute, 117924 Moscow, Russia.}
\author{Holger J Kaufmann and Ekhard K H Salje} 
\address{Interdisciplinary Research Centre in Superconductivity,
University of Cambridge, Madingley Road, Cambridge CB3 0HE, UK and
Department of Earth Sciences, University of Cambridge,
Downing Street, Cambridge CB2 3EQ, UK.}
\author{Yoad Yagil\cite{yoadaddress}}
\address{Interdisciplinary Research Centre in Superconductivity,
University of Cambridge, Madingley Road, Cambridge CB3 0HE, UK}
\maketitle
\begin{tighten}
\begin{abstract}
The normal state optical response of La$_{2-x}$Sr$_x$CuO$_4$ is 
found to be consistent
with a simple multi--component model, based on free carriers with 
strong electron--phonon interaction, localized polaronic states near 
$0.15$ eV and a mid--infrared band at $0.5$ eV. Normal state reflectance 
and absorbance of La$_{1.83}$Sr$_{0.17}$CuO$_4$ are investigated and 
their temperature dependence is explained.
Both, the ac and dc response are recovered and the quasi--linear behavior 
of the optical scattering rate up to $3000- 4000$ cm$^{-1}$ 
is found to be consistent with strong electron--phonon interaction, which
also accounts for the value of ${\rm T_c}$.
Although not strictly applicable in the superconducting state, 
our simple model accounts for the observed penetration depth 
and the optical response below ${\rm T_c}$ can be recovered by introducing
a small amount of additional carriers. Our findings suggest that the optical
response of La$_{2-x}$Sr$_x$CuO$_4$ could be explained both, in the
normal and superconducting state, by a simple
multi--fluid model with strong electron--phonon interaction if 
the gap symmetry and the temperature dependence of the $0.5$ eV MIR band
are adequately taken into account.
\end{abstract}
\end{tighten}
\pacs{74.20.De, 74.25Fy, 74.25Gz, 74.72Dn}

\section{Introduction}
Experimental studies in recent years have revealed a variety of
anomalous optical properties of high temperature superconductors (HTSC),
both in the normal and superconducting state, which are still not
fully understood theoretically.
The optical conductivity in the mid- and near--infrared 
(MIR,NIR) regimes in the metallic phase of these materials is unusually
high.
The temperature dependence is extremely weak and the superconducting
transition can hardly be observed at far--infrared (FIR)
frequencies.
In addition, powder absorbance measurements \cite{dewing,ruscher,yoad1} reveal
an unusual temperature dependence of the integrated absorbance in the
mid--infrared: 
An increase in the integrated absorbance upon cooling up to T$_{\rm c}$ is 
followed by a sharp slop change at T$_{\rm c}$, and a saturation or even a 
decrease at lower temperatures. Such behavior can neither be explained by a 
normal Fermi--liquid approach nor by conventional strong--coupling theory. 
\cite{nam}

To explain the optical response of HTSC two main approaches are considered: 
single- and multi--component models.
In single--component models the complex optical conductivity 
$\sigma(\omega)$ is written by means of a 
Drude term with both relaxation time $\tau$ and effective electron mass 
$m^*/m$ being functions of the photon energy
$\hbar\omega$,\cite{sulewski}
\begin{equation}\label{extendeddrude}
\sigma(\omega) = \frac{\omega^{2}_{\rm p}}{4\pi}\,\frac{1}{
1/\tau(\omega) - {\rm i}\omega\,m^*(\omega)/m}\quad,
\end{equation}
while multi--fluid models \cite{timusk} generally include a number of Lorentz 
oscillators, peaked at non--zero frequencies. I.e.\ 
localized states which dominate at MIR frequencies
are superimposed on a free carrier contribution in multi--component models.
Experimental data on the optical conductivity of optimal doped HTSC 
compounds cannot distinguish between the above approaches since no clear 
structures are resolved in these data. In the case of underdoped compounds 
more structure is observed, suggesting that multi--component descriptions are 
more adequate. \cite{yoad1}

A vastly enhanced MIR response of chemically doped parent compounds
of HTSC, both n type and p type, is also observed in the isolating state.
\cite{calvani1,calvani2}
Here the optical response indicates the existence of self-localized carriers:
small polarons. The MIR band, which is normally absent in stoichiometric 
samples, can be interpreted as a
composition of overtones of local modes created by the self--trapping
of extra charges injected into the lattice.
In addition, absorbance measurements of various Sr doped 
La$_{2-x}$Sr$_x$CuO$_4$ 
(LSCO) compounds demonstrated that localized states are present near $0.15$ 
eV and have a significant oscillator strength even at high doping 
levels.\cite{yoad1} 
These results show that multi--component approaches should be 
considered in the under- and optimal doped regime, and even in slightly 
overdoped compounds.

So far multi--component approaches have used a simple Drude term to 
model the free carriers contribution to the optical response.
However, there is considerable evidence for strong electron--phonon coupling 
in HTSC, and it is now widely accepted that the simple Fermi liquid
quasiparticle description is not even applicable to the normal state
metallic phase of these materials.
Indeed, while the MIR spectra depend only weakly on
temperature, the free carrier scattering rate, dominated by the
electron--phonon interaction, is strongly temperature dependent and
recent studies on YBa$_2$Cu$_3$O$_{7-\delta}$ (YBCO)\cite{yoad2}
and LSCO\cite {yoad1} show that to model this behavior
the simple Drude term for the free carriers must be modified to contain
complicated frequency and temperature dependencies which yet have to be
explained theoretically.

In this paper we will advance the multi--fluid approach by applying 
strong--coupling theory to determine the free carriers contribution to the
optical response.
As a model system we chose LSCO which is highly qualified for studying 
the optical properties of HTSC since it has no Cu--O
chains and low c--axis conductivity. Analysis of absorbance data of
under- and optimal doped compounds is straightforward using effective
medium approach \cite{yoad1,yoad2}, since the c--axis conductivity is non
metallic. For slightly overdoped LSCO,
discussed here, c--axis conductivity might effect the FIR
response but not the MIR and NIR spectra.

We will show that the normal
state response of LSCO compounds is well described by a simple 
multi--component approach based on three main contributions: (a) free carriers 
response with strong electron-phonon interaction; 
(b) localized polaronic states near $0.15$ eV; 
(c) mid--infrared band near $0.5$ eV. 
With this model one can account for the optical response and its temperature 
dependence, the dc resistivity, and the value of the 
superconducting transition temperature ${\rm T_c}$ based on normal
electron--phonon interaction pairing.

Although a variety of simplifying assumptions make our model
strictly applicable only in the normal state, it is instructive to
apply it to the superconducting state as well. 
The predicted London penetration depth is in agreement with measured data 
and the optical response below ${\rm T_c}$ can be accounted for if a small 
amount of additional charge carriers, ``hidden'' in the normal state, is 
introduced. As the most likely source of these carriers we identified the 
$0.5$ eV MIR band. This is in agreement with observed behavior of the
total absorbance and would suggest a bipolaronic origin of this band,
as proposed by Alexandrov {\it et al} (Ref.\ \onlinecite{alexbipol}).
Furthermore, an investigation of the FIR regime below ${\rm T_c}$ suggests 
that the gap--symmetry in LSCO is more complicated than simple s--wave,
in agreement with earlier results (See for example H.\ S.\ Somal
{\it et al} (Ref.\ \onlinecite{somal})
and references therein).
Our findings indicate that the optical response of LSCO, and possibly
of other HTSC, could be understood both, above and below T$_{\rm c}$,
within the framework of strong electron--phonon coupling theory
if more realistic models for the $0.5$ eV MIR band and the gap symmetry 
are used.

\section{Model}\label{lscomodel}

We model the optical response within a three liquid model, consisting of
two bands of localized charge carriers near $0.15$ eV and $0.5$ eV,
respectively, and a free carrier contribution derived within the classical
strong--coupling theory of superconductivity.
If we disregard the phonon contribution in the FIR regime,
the total dielectric function can be written as
\begin{equation}
\label{epsi}\label{threeliquidepsi}
\epsilon = \epsilon_{\infty} + \epsilon_{\rm E} + \epsilon_{\rm be_1} +
\epsilon_{\rm be_2}\;.
\end{equation}
Here, $\epsilon_{\rm be_1}$ and $\epsilon_{\rm be_2}$ represent the 
contributions of the two MIR bands, $0.15$ eV and $0.5$ eV, 
respectively, and $\epsilon_{\rm E}$ is the free carriers contribution.

In the normal state, with which we are mainly concerned here, the omission of 
phonons from the dielectric function has no substantial effect. 
We note however, that their omission magnifies the impact of the 
superconducting transition on the optical response below ${\rm T_c}$. 
In order to include phonons one should consider their temperature 
dependent line width, energy shift, and renormalization below ${\rm T_c}$
which are out of the scope of the present paper. 

The $0.15$ eV band has been experimentally identified to be of polaronic
origin \cite{yoad1,falck} since its optical activation energy is 3-5
times larger than the thermal one.
The optical response of polarons depends on their size and on the
distribution of binding energies, for example due to disorder, and might
be well approximated by a Lorentzian dielectric function. We therefore
model the $0.15$ eV band with a temperature dependent Lorentz oscillator.
It should be noted, however, that this approximation might be
over-stretched in the FIR regime, especially for temperatures
below ${\rm T_ c}$. According to the experimental results of Falck
{\it et al} (Ref.\ \onlinecite{falck}) 
the polarons are thermally delocalized, hence we
consider thermally activated charge transfer from the $0.15$ eV band to
free charge carriers.

The origin of the second MIR band, near $0.5$ eV, is yet unknown.
One possibility is the dissociation of bipolarons which do or do not form a
conduction band. Experimentally it was shown that this band is only weakly
temperature dependent and that the total absorbance is modified below the
superconducting transition. In this work we ignore this temperature 
dependence since it is hardly observed in reflectance data, and approximate 
the $0.5$ eV band with a temperature independent Lorentzian. Although this 
simplification is well justified in the normal state, there is evidence that 
it does not hold in the superconducting state.

The Lorentz dielectric functions for the two MIR bands have the form
\begin{equation}
\epsilon_{{\rm be}_j}(\omega) =
\frac{\omega^2_{{\rm p}_{({\rm be}_j)}}}{\omega^2_{{\rm e}_j}
-\omega^2-
{\rm i}\omega\gamma_{{\rm e}_j}}\:,
\;\; j=1,2\;.
\end{equation}
If we neglect any difference in effective mass between the polaronic and free
states, the free carriers plasma frequency and the oscillator
strength of the polaronic Lorentzian obey the sum rule
\begin{equation}
\omega^{2}_{\rm p_E}({\rm T})+
\omega^{2}_{\rm p_{(be_1)}}({\rm T})
= {\rm const.\ }\;
\end{equation} and
the temperature dependent spectral weight of the $0.15$ eV band is
given by
\begin{mathletters}
\begin{equation}\label{omegap}
\omega^{2}_{\rm p_{(be_1)}}({\rm T}) =
\omega^{2}_{\rm p_{(be_1)}}(0)\,\left[1-\exp\left(
-\frac{\rm T_0}{\rm T}\right)\right]\;,
\end{equation}
while the plasma frequency of the free carriers is
\begin{equation}\label{omegae}
\omega^{2}_{\rm p_E}({\rm T}) = \omega^{2}_{\rm p_E}(0) +
\omega^{2}_{\rm p_{(be_1)}}(0
)\exp\left(-\frac{\rm T_0}{\rm T}\right)\;.
\end{equation}\end{mathletters}
Here, ${\rm T_0}$ denotes a thermal activation energy.

The free carriers contribution to the optical response
is calculated by applying conventional strong--coupling
theory. Thus, we assume that the conduction electrons in the Cu--O planes
behave like in a conventional electron--phonon superconductor, which can be
treated by Migdal--Eliashberg theory, and for simplicity we assume 
s--wave pairing.
The later assumption is of course an oversimplification which might effect
the FIR spectra below ${\rm T_c}$.
In this strong coupling extension of BCS theory expressions for the optical 
conductivity, disregarding vertex corrections for the electron--phonon 
interaction, are well established now. \cite{nam,Lee,dolgov,akis} 
They require the solutions of the Eliashberg
equations. \cite{eliashberg}
For a two--dimensional isotropic system with cylindrical Fermi
surface the Eliashberg equations for the renormalized
gap parameter $\tilde{\Delta}$ and energy $\tilde{\epsilon}$
can be written in the following form \cite{dolgov}
\begin{mathletters}
\begin{equation}
\label{elias1}
\tilde{\epsilon}(\epsilon) = 1 -
\int\limits_{-\infty}^{\infty} {\rm d}\omega
\int\limits_{0}^{\infty} {\rm d} \Omega \; \alpha^2F(\Omega)
I(\epsilon + {\rm i}\delta, \Omega, \omega
) \; {\rm Re} \frac{  \tilde{\epsilon}(\omega) }{
\sqrt{ \tilde{\epsilon}^2(\omega)
 - \tilde{\Delta}^2(\omega)
 } } \quad,
\end{equation}
\begin{eqnarray}
\label{elias2}
\tilde{\Delta}(\epsilon) & = & \nonumber
- \int\limits_{-\infty}^{\infty} {\rm d}\omega \int\limits_{0}^{\infty} 
{\rm d} \Omega \; \alpha^2F(\Omega)
I(\epsilon + {\rm i}\delta, \Omega,\omega) \; {\rm Re} \frac{
\tilde{\Delta}(\omega) }{
\sqrt{ \tilde{\epsilon}^2(\omega) - \tilde{\Delta}^2(\omega) } }
\\  &&
- \frac{1}{2}\mu^*(\omega_c) \int\limits_{-\omega_c}^{\omega_c}
{\rm d}\omega \; {\rm tanh}\frac{ \omega}{2 {\rm T}} \;
{\rm Re} \frac{ \tilde{\Delta}(\omega) }{\sqrt{ \tilde{\epsilon}^2(\omega)
- \tilde{\Delta}^2(\omega) } }
\quad,
\end{eqnarray}\end{mathletters}
where
\begin{equation}
I(\epsilon + {\rm i}\delta, \Omega, \omega) =
\frac{ N(\Omega) + 1 - f(\omega) }{ \epsilon +{\rm i}\delta -\Omega-\omega} +
\frac{ N(\Omega) + f(\omega) }{ \epsilon +{\rm i}\delta +\Omega-\omega}
\quad,\end{equation}
$\mu^*$ is the Coulomb
pseudo-potential and $\omega_c$ is the frequency cutoff. $N(\Omega)$ and
$f(\omega)$ are Bose and Fermi distribution functions.
As usual, the impurities are assumed to be non--magnetic here. Consequently,
impurity scattering does not appear in the isotropic equations (\ref{elias1}) 
and (\ref{elias2}).

Using the standard theory of electromagnetic response function, the
optical conductivity can subsequently be calculated in the local (London)
limit. Therefore, the free carriers contribution to the
optical conductivity is given by \cite{Lee}
\begin{eqnarray}
\label{sigma}
\sigma_{\rm E}(\omega,{\rm T}) = 
\frac{\omega^2_{\rm p_{\rm E}}}{ 2\pi\,\omega} \;
\int\limits_{-\infty}^{\infty} {\rm d}\epsilon \;
\left[ {\rm tanh}\frac{\epsilon}{2k_B {\rm T}} M(\epsilon,\omega)
\left( g(\epsilon)g(\epsilon+\omega) + h(\epsilon)h(\epsilon+ \omega) + \pi^2
\right)\right.
\\ 
\nonumber
-{\rm tanh}\frac{\epsilon+\omega}{2k_B {\rm T}} M^* (\epsilon,\omega)\left(
g^* (\epsilon)g^* (\epsilon+\omega) + h^* (\epsilon)h^* (\epsilon+ \omega) +
\pi^2 \right)
\\
\nonumber
+
{\rm tanh}\frac{\epsilon+\omega}{2k_B {\rm T}}
 L(\epsilon,\omega) \left(
g^* (\epsilon)g(\epsilon+\omega) + h^* (\epsilon)h(\epsilon+ \omega) + \pi^2
\right)
\\ \nonumber
\left. - {\rm tanh}\frac{\epsilon}{2k_B {\rm T}}
 L(\epsilon,\omega) \left(
 g^* (\epsilon)g(\epsilon+\omega) + h^* (\epsilon)h(\epsilon+ \omega) + \pi^2
 \right)\right]\quad,
\end{eqnarray}
where
\begin{mathletters}
\begin{eqnarray}
g(\epsilon) &=& \frac{ -\pi\, \tilde{\epsilon}(\epsilon)}{
\sqrt{ \tilde{\Delta}^2(\epsilon) - \tilde{\epsilon}(\epsilon)}}\quad,\\
h(\epsilon) &=& \frac{ -\pi\, \tilde{\Delta}(\epsilon)}{
\sqrt{ \tilde{\Delta}^2(\epsilon) - \tilde{\epsilon}(\epsilon)}}\quad.
\end{eqnarray}\end{mathletters}
The functions
\begin{mathletters}
\begin{equation}
M(\epsilon,\omega) = \left[
\sqrt{ \tilde{\Delta}^2(\epsilon+\omega) - \tilde{\epsilon}^2
(\epsilon+\omega)} +
\sqrt{ \tilde{\Delta}^2(\epsilon) - \tilde{\epsilon}^2
(\epsilon)} + \gamma_{\rm imp} \right]^{-1}\quad,
\end{equation}
and
\begin{equation}
L(\epsilon,\omega) = \left[
\sqrt{ \tilde{\Delta}^2(\epsilon+\omega) - \tilde{\epsilon}^2
(\epsilon+\omega)} +
\sqrt{ \tilde{\Delta}^{*2}(\epsilon) - \tilde{\epsilon}^{*2}
(\epsilon)} + \gamma_{\rm imp} \right]^{-1}\quad,
\end{equation}\end{mathletters}
include normal impurity scattering effects in Born approximation,
$\gamma_{\rm imp}$ is an average scattering rate, and $\omega^2_{\rm p_E}$ 
denotes the effective plasma frequency of the free carriers in the 
Cu--O planes. The free carriers contribution to the total dielectric
function (\ref{epsi}) is then given by $\epsilon_{\rm E}(\omega)= 
4\pi{\rm i}\sigma_{\rm E}(\omega,{\rm T})/\omega$.

The strong--coupling theory is no first principle theory but requires definite 
assumptions on the Eliashberg function, $\alpha^2F(\omega)$,
and needs coupling parameters as an input. 
Here we started from published phonon density of state \cite{renker} 
but found it incompatible with the optical and the dc data since the 
scattering rate at low frequencies is too large. Better agreement is achieved 
when $\alpha(\omega)$ is nonuniform and increases at high frequencies. 
This is also consistent with tunneling data \cite{bulaevskii,Deutscher} 
where high energy phonons seem to have stronger electron--phonon interaction
coefficient. For the sake of simplicity we chose a power law behavior: 
$\alpha^2(\omega) = (\omega/\omega_0)^{\eta}$
with $\eta > 0$. The phonon spectrum $F(\omega)$ is estimated by a set of 
quadratic Lorentzians and a low frequency cutoff function to account for the 
low frequency tail which otherwise dominates the electron--phonon interaction 
constant  
\begin{equation}
\lambda = 2\int\limits_{0}^{\infty}\!\frac{{\rm d}\omega}{\omega}\:
\alpha^{2}(\omega){\rm F}(\omega)\quad.
\end{equation}
The resulting 
Eliashberg function is shown in Fig.\ \ref{phonondos} and the parameters 
used to define $\alpha^{2}(\omega){\rm F}(\omega)$ are listed in Table 
\ref{elfidata}.
The overall coupling coefficient $\lambda$ was adjusted to give the 
correct value of T$_{\rm c}$. To determine the superconducting transition
temperature for our choice of parameters we used McMillan's equation. 
\cite{allen} The resulting ${\rm T_c}$ is in good agreement with
the transition temperature derived directly from Migdal--Eliashberg theory by
calculating the London penetration depth $\lambda_{\rm L}({\rm T})$.
In the superconducting state, for ${\rm T}$ close to ${\rm T_c}$,
one expects $\lambda^{-2}_{\rm L}({\rm T})\propto {\rm T_c}-{\rm T}$.
Since
\begin{equation}\label{londonsigma}
\lambda_{\rm L}^{-2}({\rm T})=
\lim\limits_{\omega\rightarrow 0}\,4\pi\/\omega\,{\rm Im}\,
\sigma(\omega,{\rm T})\;,
\end{equation}
the transition temperature can be easily determined once $\sigma_{\rm E}(
\omega)$ is calculated.
Similarly, the impurity scattering rate 
$\gamma_{\rm imp}$ is found by the measured dc resistivity, and the 
strength of localized states is in general agreement with those found 
in absorbance measurements. \cite{yoad1}

We have performed various model calculations using different values for
the electron--phonon coupling strength $\lambda$, ranging from $0.7-1.4$.
These calculations show that it is necessary to assume $\lambda\approx 1$ 
to achieve reasonable agreement with both, tunneling data and
the optical and dc data.
In general, we found that the different kinds of data investigated here
put severe 
constraints on possible choices of the fitting parameters and
we note that although the number of model parameters is rather large,
no significant deviation from the parameters listed below (Table 
\ref{fitparameters1}) are possible.

\section{Results and Discussion}
Using the model described in Sec.\ \ref{lscomodel}
we have calculated the optical reflectance and dc resistivity of  
slightly overdoped La$_{2-x}$Sr$_x$CuO$_4$, $x=0.17$.
The calculated curves were then fitted to the experimental data of Gao
{\it et al} (Ref.\ \onlinecite{gao}) at  ${\rm T}=200$ K with the addition 
of a high frequency Lorentzian at $1.5$ eV. The resulting fit parameters are
listed in Table \ref{fitparameters1}.
Figure \ref{reflnormal1} shows the calculated normal state reflectance,
\begin{equation}
R(\omega) = \left|(\epsilon^{1/2}(\omega) -1)/(\epsilon^{1/2}(\omega) + 1)
\right|^2\;,
\end{equation}
together with the measured ones at three different temperatures
and an additional calculated curve at $40$ K.
At low frequencies ( $<1000$ cm$^{-1}$)
the optical response is dominated by the free carriers and is therefore
temperature dependent. At higher frequencies the temperature
dependence is largely suppressed since the spectral weights of the two
localized states become appreciable and since the opposite trends of the
$0.15$ eV band and the free carrier part result in partial compensation.
At higher frequencies ( $> 3000$ cm$^{-1}$) the calculated spectrum is
dominated by the $0.5$ eV band and is practically temperature independent.
The thermal activation energy of the $0.15$ eV band, ${\rm T}_0 = 500$
cm$^{-1}$, is by roughly a factor of 4 smaller than the optical energy,
indicating self--localized charge carriers: while the lattice is frozen for
the fast optical transitions (Frank--Condon principle), in thermal excitations
it has time to relax and the polaronic binding energy is consequently
reduced. The activation energy ${\rm T}_0 = 500$ cm$^{-1}$
is very close to the polaronic energies found in insulating parent compounds
of HTSC.\cite{calvani1,calvani2} Moreover, photoexcited carrier relaxation
measurements by Mihailovic {\it et al} provide evidence for localized
polaronic states at similar energies in both, LSCO and
YBCO.\cite{mihail}

Another quantity of interest is the dielectric loss function, ${\rm Im}(-1/
\epsilon(\omega))$. Experimentally the loss function of HTSC is found 
to follow a quadratic behavior,
${\rm Im}(-1/\epsilon(\omega))\propto \omega^2$, over almost the
entire frequency range up to the plasma edge, where
it exhibits a (first) peak and subsequently decreases.\cite{bozovic} 
The height of the peak is temperature dependent, while its position and  
the curvature of ${\rm Im}(-1/\epsilon(\omega))$ strongly depend on  
$\epsilon_\infty$.
We have calculated the dielectric loss function using a variety
of different model parameters, confirming the high sensitivity of the peak 
position to the choice of $\epsilon_\infty$. 
For optimally doped ($x=0.15$) LSCO the
position of the maximum has been reported
to be approximately $6500$ cm$^{-1}$.\cite{bozovic}  
Figure \ref{lossfct} shows the dielectric loss function for
La$_{1.83}$Sr$_{0.17}$CuO$_4$ at $200$ K as calculated within our
multi--liquid model using the parameters from Table \ref{fitparameters1}.
The behavior in the FIR and MIR regimes is quadratic to  good approximation
and the curve shows a maximum at $6320$ cm$^{-1}$, close to the
experimentally observed value for optimal doping.
This confirms our choice of $\epsilon_\infty=4.43$.

With the parameters obtained from the fit of the reflectance data
the absorption coefficient
$ \alpha(\omega)=4\,\pi\,\omega\,{\rm Im}(\epsilon^{1/2}(\omega)) $
can be calculated and compared to measured optical
absorbance data. Analysis of absorbance data must take into account 
grain--size effects and the response of the host material. Therefore, the
major disadvantage of this approach is the difficulty in converting qualitative 
results into quantitative ones. However, it has been shown recently
that absorbance measurements at MIR frequencies can be analyzed
quantitatively within the framework of 
the effective medium approach.\cite{yoad2}
Figure \ref{absexp} shows normal state 
powder absorbance data of La$_{1.83}$Sr$_{0.17}$CuO$_4$ 
from Ref.\ \onlinecite{yoad1}. 
In Fig.\ \ref{abscalc} the corresponding calculated curves are shown.
In agreement with the experimental results, $\alpha(0{\rm K}) -
\alpha(300{\rm K}) \approx 10^4$ cm$^{-1}$ near $0.15$ eV. The overall
behavior of the absorbance is well resembled in Fig.\ \ref{abscalc}:
All curves show the characteristic maximum close to the $0.5$ eV band and
the absorption coefficient decreases with increasing temperature for 
frequencies up to $3000$ cm$^{-1}$. 

The normal state response can also be described within the framework of an
extended Drude model where the relaxation time $\tau$ and
the effective mass $m^*$ are allowed to vary with
frequency.
Writing the normal state optical conductivity derived within our 
three--liquid model in a generalized Drude formula, Eq.\ (\ref{extendeddrude}),
the scattering rate $1/\tau$ and the effective mass $m^*/m$ can be calculated.
They are shown in Fig.\ \ref{extendeddrudefig}, together with 
the ''optical scattering rate``,
\begin{equation}
\frac{1}{\tau^*} = \omega\,\frac{{\rm Re}(\sigma)}{{\rm Im}(\sigma)}
=\frac{m}{m^*}\,\frac{1}{\tau}
\;.
\end{equation}
The quasi-linear behavior of $1/\tau^*$ up to $3000-4000$ cm$^{-1}$  
was one of the driving forces for assuming electron-electron interaction
(marginal \cite{marginal} and nested \cite{nested} Fermi Liquids for example)
rather than electron-phonon one since the phonon spectrum does not exceed
$1000$ cm$^{-1}$. As already mentioned by Shulga {\it et al}
(Ref.\ \onlinecite{dolunpublished}) this argument is
valid for $1/\tau$ rather than $1/\tau^*$ which indeed becomes frequency
independent above the phonon spectrum. Therefore the multi--component model
with strong electron--phonon interaction supports one of the major
properties of HTSC:
the quasi--linear response of the optical scattering rate
up to energies much higher than the phonon spectrum. It is
important to note here that obtaining  $1/\tau^*$ experimentally is 
somewhat delicate since this quantity is quite sensitive to the value of 
$\epsilon_{\infty}$.

The general behavior of the dc resistivity also support the above picture,
namely the co--existence of free and localized charge carriers and the thermal
excitations of the later.
Temperature dependent resistivity measurements
are done at constant pressure, and the correction to constant volume reveal
deviation from quasi--linear behavior where the resistivity data tend to
saturate or follow a lower slope at high energies. This change of slope is
also noticed in the calculated resistivity, where the larger number of free
carriers reduces the slope of the $\rho({\rm T})$ curve, Fig.\
\ref{resistisity}.
Another manifestation of the above behavior is the doping dependence of
${\rm d}\rho/{\rm dT}$. Uchida has measured the temperature
dependence of several LSCO compounds up to $950$ K.\cite{uchida} 
A clear deviation from linear temperature dependence is observed at
high temperatures for the $x=0.1$ and $x=0.12$
compounds even in the constant pressure data. Moreover, a rough estimate of the
ratios between the slopes $[{\rm d}\rho/{\rm dT}]_x/[{\rm d}
\rho/{\rm dT}]_{0.1}$ yields  $0.1$, $0.2$, $0.36$,
$0.58$ for $x=0.3$, $0.2$, $0.15$, $0.12$, respectively.
Assuming that $\lambda$ does not change much for these compounds one expects
${\rm d}\rho/{\rm dT} \propto 1/\omega_{\rm p}^2 \propto 1/(x-x_0)$, where
$x_0$ is the total number of localized carriers per Cu atom.
The above ratios yield $x_0=0.075$ in close agreement with the estimated
spectral weight from absorbance measurements.\cite{yoad1}

Once the parameters of our model are determined as described above,
the optical response of LSCO can
be calculated for arbitrary temperatures, above and below
${\rm T_c} = 37.4$ K.
Therefore, we will also apply the model to the superconducting state to 
gain an insight in how it could and should be altered to be applicable
in both, the normal and superconducting state.
 
The first quantity of interest here is the London penetration depth. 
According to Eq.\ (\ref{londonsigma}), its zero--temperature value, 
$\lambda_{\rm L}(0)$, provides a measure
for the free carriers plasma frequency $\omega_{\rm p_E}(0)$,
since the zero--frequency limit of the conductivity is entirely due to the
free carriers response.
Figure \ref{london1} shows the normalized
inverse penetration depth, $\lambda_{\rm L}^2(0)/\lambda^2_{\rm L}({\rm T})$,
as predicted by our three--liquid model, using the parameters from Table
\ref{fitparameters1}.
At low temperatures the phenomenological relation, 
\begin{equation}
\lambda_{\rm L}({\rm T})\approx
\lambda_{\rm L}(0)\left[1-\left(\frac{\rm T}{\rm T_c}
\right)^4\,\right]^{-1/2}\;,
\end{equation}
is well resembled and we can use this expression
to extrapolate $\lambda_{\rm L}({\rm T})$ to zero temperature.
This yields $\lambda_{\rm L}(0) = 2579$ \AA.
Experimentally it is difficult to obtain the absolute value of the 
penetration depth, especially for thin films. 
However, it is generally accepted that $\lambda_{\rm L}(0) > 
2300$ {\AA} for optimally doped LSCO\cite{kossler,aeppli,uemera}.
Using muon spin relaxation rate measurements Aeppli {\it et al}
found $\lambda_{\rm L}(0)\approx 2500$ {\AA}, close to the value predicted by
our multi--liquid approach.

It has been observed earlier for various HTSC that the integrated absorbance 
exhibits an unusual temperature dependence.\cite{dewing,ruscher,yoad1}
In the normal state the integrated absorbance increases upon cooling.
At the superconducting transition temperature a sudden slope change occurs,
followed by a saturation or even a decrease of the integrated absorbance
upon further cooling. Figure \ref{integrabs} shows the measured\cite{yoad1} 
absorbance of 0.17 Sr LSCO integrated over the frequency range 
$1600$ cm$^{-1} < \omega < 4500$ cm$^{-1}$, normalized to unity at 40K. 
At low temperatures the experimental uncertainties are fairly large, but a 
saturation effect below T$_{\rm c}$ can clearly be seen. 
This indicates that spectral weight is shifted from the $0.5$ eV band into 
the $\delta$--peak at zero--frequency upon cooling below ${\rm T_c}$.
Within ordinary strong--coupling theory such behavior
cannot be understood since $\omega \sim 0.5\;{\rm eV} \gg 2\Delta$, and
it should not be possible to reproduce this behavior within our
multi--fluid model where the $0.5$ eV MIR band has been assumed to be 
temperature--independent.
Indeed, a saturation of the calculated total absorption coefficient 
can only be observed if the range of integration is extended
well into the FIR regime, as shown in Fig.\ \ref{intabs}.
This indicates that 
the observation of the superconducting transition
at NIR frequencies is not merely a result of 
Kramers--Kronig relations. The fact that
an explanation of the observed change in integrated MIR absorbance
is beyond our model suggests instead some temperature dependence of the $0.5$ eV
band and supports the approach of Alexandrov {\it et al} (Ref.\
\onlinecite{alexbipol}).

Another discrepancy between the measured and calculated optical response
can be seen in the reflectance below ${\rm T_c}$.
While the normal state fit curves practically coincide with the experimental
reflectance data over a wide range of frequencies (see Fig.\ \ref{reflnormal1}),
in the superconducting state the measured reflectance from Gao {\it et al}
is slightly underestimated by our model. 
This could be simply due to the fact that the simplifying assumption of equal 
effective mass for the free and localized carriers breaks down below 
${\rm T_c}$.
Alternatively, the discrepancy can be accounted for by introducing 
additional free carriers. Figure \ref{reflsuper2} shows the measured
reflectance at $10$ K together with the prediction of our model. Here,
a slightly modified model, with an additional small Drude term, 
with $\omega_{\rm p_D}=2000$ cm$^{-1}$ and $\gamma_{\rm D}
=200$ cm$^{-1}$, has been included. The close resemblance between the 
measured curve in Fig.\ \ref{reflsuper2} with this modified multi--liquid
model indicates the existence of a ``hidden'' reservoir of charge carriers
which has to be taken into account in the superconducting state. 
There is increasing evidence that the gap symmetry in LSCO is more complicated
than simple s--wave.\cite{somal} In the case of d--wave pairing, the 
``hidden'' carriers could be explained by the contribution of resonant
scattering.
Another likely source for these additional carriers is the second MIR band,
near $0.5$ eV, which was taken to be temperature independent in our simple 
multi--fluid approach. 
Indeed, by increasing the free carriers plasma frequency slightly, to
$\omega_{\rm p_E}(0)=10591$ cm$^{-1}$, and decreasing the oscillator strength
of the second Lorentzian accordingly, 
to $\omega_{\rm p_{(be2)}}=5748$ cm$^{-1}$,
the MIR and NIR reflectance in the superconducting state can be 
easily accounted for. 
The consequent shift in spectral weight and the position of the second MIR band
would then suggest a bipolaronic origin of this band.
Alexandrov {\it et al} (Ref.\ \onlinecite{alexbipol}) 
have shown that the unusual behavior of the integrated absorbance
can be explained by Bose--condensation of bipolarons which are also 
responsible for the second MIR band. The comparison of our model with the 
measured reflectance in the superconducting state seems to confirm this 
picture.
In Fig.\ \ref{bipol} the reflectance curve at $10$ K calculated
with the modified free carriers plasma frequency and oscillator strength
of the second Lorentzian is shown for frequencies below $2500$ cm$^{-1}$
together with our original model and the measured reflectance.
Above $1500$ cm$^{-1}$ the modified spectral
weight provides better agreement with the experimental data.
Between $100-600$ cm$^{-1}$ our model predictions can be seen to deviate from
the experimental curve, predicting lower absorbance in this frequency regime.
This is due to the simplifying assumption of s--wave pairing for the 
free carriers. As mentioned above, the gap symmetry 
in LSCO is probably more complicated and must be taken into account to 
describe the FIR regime, since its shape leaves long tails which can easily 
alter the optical response up to $1000$ cm$^{-1}$. In addition, 
the omission of phonon modes and the simplified model
for the polaronic dielectric function also affects the FIR spectra,
especially below ${\rm T_c}$. 

The correspondence of the normal state optical response of LSCO with the simple
model consisting of a band of localized charge carriers, a 
temperature independent MIR band near $0.5$ eV, and a free carrier 
contribution treated within ordinary strong coupling
theory suggests the following physical picture for LSCO and possibly for
other HTSC:
The charge carriers introduced by doping in HTSC parent compounds are
redistributed and form polaronic bound states. With increased doping levels,
these states fill up to the overcrowding limit where additional bound states
start to overlap existing ones. Additional doping results in the appearance
of charge carriers which form a Fermi--liquid with strong electron--phonon
interaction. 
The comparison of this model with experimental data in the superconducting
state suggests further that (a) the gap symmetry in LSCO is more
complicated than simple s--wave and (b) the broad MIR band at $0.5$ eV might be
of bipolaronic origin.

\section{Summary}
Our model, based on a free carrier response with strong electron--phonon 
interaction, localized polaronic states near $0.15$ eV and a 
temperature independent mid--infrared band
near $0.5$ eV reveals the following results: 
The normal state optical response of LSCO from the far--infrared to the  
near--infrared is consistent with our multi--liquid approach. Assuming strong 
electron--phonon coupling both the ac and dc response are recovered, the 
later up to very high temperatures.  
The quasi--linear behavior of the optical scattering rate for frequencies up 
to $3000-4000$ cm$^{-1}$ is also consistent with this approach and in agreement
with previous measurements \cite{yoad1} we find the doping dependence
${\rm d}\rho/{\rm dT} \propto
1/(x-x_0)$, where $x_0=0.075$.
The polaronic origin of the $0.15$ eV band is confirmed. The temperature
dependence of the optical response, both reflectance and absorbance, is
explained and our model naturally accounts for the value of the 
superconducting transition temperature.

In the superconducting state the absolute value of the London penetration
depth is recovered. A comparison of the calculated reflectance 
and integrated absorbance with measured data indicates that the simplifying
assumption of a temperature independent mid--infrared band at $0.5$ eV 
breaks down in the superconducting state.
Model calculations would then suggest a bipolaronic origin of this band, 
in agreement with earlier results.\cite{alexbipol}
Our model used the simplifying assumption of s--wave pairing.
The far--infrared spectra suggest that the symmetry of the gap
parameter is more complicated for LSCO. 
All this indicates that a multi--component approach can be extended
to the superconducting state by more realistic models for both, the $0.5$
eV mid--infrared band and the gap symmetry. 
Hence, the optical response of LSCO, both
above and below T$_{\rm c}$, could be understood within the framework
of strong electron--phonon coupling theory which is also the driving mechanism
for pairing if the localized states which dominate the mid--infrared
response are adequately taken into account.

\section*{Acknowledgment}
We gratefully acknowledge David Tanner for providing us with his  
reflectance data.
One of us (H.\,J.\,K.) would like to thank V.\ V.\ Kabanov for 
many helpful discussions and
O.\,V.\,D.\ would like to thank 
the Royal Society and the Department of
Earth Sciences, University of Cambridge for  support and hospitality
during the early stages of this work.

\newpage
\begin{tighten}
\begin{table}
\caption{Parameters used to define the Eliashberg function $\alpha^2F(\omega)$.
$\omega_{min}=60$ cm$^{-1}$, $\omega_\alpha = 1000$ cm$^{-1}$, $A_\alpha 
=3.8$, and $\eta = 1.9$, where: $\alpha^2(\omega) = 
A_\alpha(\omega/\omega_\alpha)^\eta$, 
$F(\omega) = \frac{2}{\pi}\,\tanh^{-1}\left((\omega/\omega_{min})^4\right)
\,\sum_i \: A_i \left( 1+(\omega - \omega_{0_i})^2
/\omega^{2}_{d_i}\right)^{-2}$}\label{elfidata}
\begin{tabular}{lrr}
$A$&$\omega_0$&$\omega_d$\\
\tableline
0.2&80&40\\
0.25&160&50\\
0.3&230&30\\
0.2&320&35\\
0.18&380&30\\
0.12&460&40\\ 
0.05&560&40\\
0.02&680&50\\
\end{tabular}\end{table}
\begin{table}
\caption{Parameters used to fit the optical and dc resistivity data of Gao
{\it et al}. 
For ${\rm T}>250$ K the measured reflectance at high
frequencies indicates a phase transformation. In order to account for this
change in the calculation of the $300$ K data we set
$\epsilon_{\infty}=3.9$ and $\omega_{{\rm p}_{(be_3)}}=9000$ cm$^{-1}$.
No further changes were needed in order to fit the low and high temperature
data.}\label{fitparameters1}
\begin{center}
\begin{tabular}{lr}
$\lambda$&1.0\\
$\mu^{*}$& 0.05\\
$\gamma_{\rm imp}$ & 100 cm$^{-1}$\\
\tableline
$\epsilon_{\infty}$ & 4.43 \\
T$_0$ & 500  cm$^{-1}$\\
$\omega_{\rm p_{E}}(0)$ & 10400 cm$^{-1}$\\
\tableline
$\omega_{{\rm p}_{({\rm be}_1)}}(0)$ & 7840 cm$^{-1}$\\
$\omega_{\rm e_1}$ & 1650 cm$^{-1}$\\
$\gamma_{\rm e_1}$ &3550 cm$^{-1}$\\
\tableline
$\omega_{{\rm p}_{({\rm be}_2)}}$ & 5834  cm$^{-1}$\\
$\omega_{\rm e_2}$ & 3740 cm$^{-1}$\\
$\gamma_{\rm e_2}$ &5056 cm$^{-1}$\\
\tableline
$\omega_{{\rm p}_{({\rm be}_3)}}$ & 6000  cm$^{-1}$\\
$\omega_{\rm e_3}$ & 11000 cm$^{-1}$\\
$\gamma_{\rm e_3}$ &5000 cm$^{-1}$\\
\end{tabular}
\end{center}
\end{table}

\begin{figure}[ht]
\caption{Eliashberg function $\alpha^2F(\omega)$ used to model the
free carriers response of La$_{1.83}$Sr$_{0.17}$CuO$_4$.
The corresponding parameters are listed in Table \ref{elfidata}.}
\label{phonondos}
\end{figure}

\begin{figure}[ht]
\caption{Measured\cite{gao} and calculated reflectance of
La$_{1.83}$Sr$_{0.17}$CuO$_4$ thick film. 
The fine structures at low frequencies (detailed scale,
lower panel) are due to infrared phonons while the overall structure is due
to strong electron--phonon interaction.}
\label{reflnormal1}
\end{figure}

\begin{figure}[ht]
\caption{Dielectric loss function ${\rm Im}(-1/\epsilon)$ as a function of
frequency for La$_{1.83}$Sr$_{0.17}$CuO$_4$ at $200$ K as predicted by our
three--liquid model.}
\label{lossfct}
\end{figure}

\begin{figure}
\caption{Measured \cite{yoad1} powder absorbance (in absorbance units)
for La$_{1.83}$Sr$_{0.17}$CuO$_4$
at various temperatures. The structure at $3700$ cm$^{-1}$ is an experimental
artifact. On lowering the temperature the optical absorbance
is enhanced over a wide frequency range.}\label{absexp}
\end{figure}

\begin{figure}
\caption{Absorption coefficient $ \alpha(\omega)=4\,\pi\,
\omega\,{\rm Im}\{\epsilon^{1/2}(\omega)\} $ for 
La$_{1.83}$Sr$_{0.17}$CuO$_4$
calculated within
the multi--liquid model with strong electron--phonon interaction using the 
parameters from Table \ref{fitparameters1}.}\label{abscalc}
\end{figure}

\begin{figure}[ht]
\caption{(top) Scattering rate $1/\tau$ and (middle) effective mass $m^*/m$ in
a generalized Drude model derived from our multi--fluid model. (bottom)
The ''optical scattering rate`` $1/\tau^*$ derived within the
multi--fluid model with strong electron--phonon interaction. While
$1/\tau$ becomes practically frequency independent above the phonon spectrum
$1/\tau^*$ shows quasi--linear behavior for frequencies up to
$3000 - 4000$ cm$^{-1}$.}
\label{extendeddrudefig}
\end{figure}

\begin{figure}[ht]
\caption{Calculated dc resistivity (solid line).
The broken line indicates the dc resistivity
without the delocalization of the $0.15$ eV states, namely with constant
number of free carriers.}
\label{resistisity}
\end{figure}

\begin{figure}[ht]
\caption{Normalized inverse London penetration depth as a function
of reduced temperature, 
calculated from the optical conductivity as predicted by the
multi--fluid model using the parameters from Table \ref{fitparameters1}.
Also shown are the phenomenological curve $1-({\rm T}/{\rm T_c})^4$ and
a straight--line fit near ${\rm T_c}=37.4$ K.}
\label{london1}
\end{figure}

\begin{figure}[ht]
\caption{Measured\cite{yoad1} integrated absorbance for
La$_{1.83}$Sr$_{0.17}$CuO$_4$, normalized to unity at 40K. 
The integration is over the spectral 
segment $1600-4500$ cm$^{-1}$.}
\label{integrabs}
\end{figure}

\begin{figure}[ht]
\caption{Calculated absorption coefficient derived within the
multi--liquid model using the parameters from Table \ref{fitparameters1},
integrated from $100 - 4500$ cm$^{-1}$.}
\label{intabs}
\end{figure}

\begin{figure}[ht]
\caption{Measured reflectance for La$_{1.83}$Sr$_{0.17}$CuO$_4$ at $10$ K
compared to our multi--fluid model, using the parameters from 
Table \ref{fitparameters1} (dash--dotted line). In contrast to the good
agreement between data and fit in the normal state (compare Fig.\ 
\ref{reflnormal1}), the measured reflectance at $10$ K is slightly 
underestimated by our model. Also shown is a modified multi--fluid model
(solid line), 
including an additional Drude term with $\omega_{\rm p_D}=2000$ cm$^{-1}$ and 
$\gamma_{\rm D}=200$ cm$^{-1}$, which better resembles the measured 
reflectance.} 
\label{reflsuper2}
\end{figure}

\begin{figure}[ht]
\caption{Comparison between the measured reflectance at $10$ K, the
multi--liquid model using the parameters from Table \ref{fitparameters1}
(dotted line), and
the model with weaker MIR response, $\omega_{\rm p_{(be2)}}=5748.1$ cm$^{-1}$
and $\omega_{\rm p_E}(0)=10591$ cm$^{-1}$ (solid line). The discrepancy
in the FIR regime is due to various simplifying model assumptions, as
explained in the text. Above $1500$ cm$^{-1}$ the measured data is well
resembled by the modified model.}
\label{bipol}
\end{figure}

\end{tighten}
\end{document}